# SITE-CITY INTERACTION THROUGH MODIFICATIONS OF SITE EFFECTS

J.F. Semblat[1], M. Kham[2], P. Guéguen[3], P.Y. Bard[3] and A.M. Duval[4]

## ABSTRACT

The analysis of seismic site effects generally disregards the influence of surface structures on the free field motion in densely urbanized areas. This paper aims at investigating this particular problem called site-city interaction especially by comparison to the "free-field" amplification process. Several evidences (experimental, analytical, numerical) of the site-city interaction phenomenon have been given in previous work (Guéguen, Bard, Semblat 2000). The influence of site city-interaction could be large for structures having eigenfrequencies close to that of the surface soil layers. Furthermore, the density of structures is also an important governing parameter of the problem.
Considering a specific site (Nice, France) where site-city interaction is supposed to be significant, we start from detailed experimental and numerical studies of seismic site effects giving both amplification levels and occuring frequencies, as well as the location of the maximum amplification areas. The influence of site-city interaction is then investigated through various numerical models considering the boundary element method. The effect of surface structures with variable urban densities is analysed to estimate the contribution of overall site-city interaction on surface motion distribution. The main goal of the paper is to estimate the influence of site-city interaction not only on amplification levels (the influence can be small), but also on the modification of both main amplification frequencies and location of the maximum amplification areas. The main conclusion of this work is to show the modifications of the free-field amplification due to site-city interaction and leading to a specific *urban field amplification*.

## Introduction

The analysis of seismic site effects can be performed by various means : experimental, analytical, numerical methods, vibratory or propagative approaches... (Bard 1985, Paolucci 1999, Semblat 2000a,b) The amplification level is mainly influenced by surface layers thickness, geometry, shear moduli as well as the wave type, source location and directivity... The geometrical and mechanical features of the medium involved in the propagation process are then essential. Considering alluvial basins located in large cities, one could have a high density of buildings along the "free" surface. The geometrical and mechanical features of the whole basin-buildings system could be different from that of the basin alone system. Furthermore, as shown in previous work (Guéguen, Bard, Oliveira 2000), the vibratory motion of surface structure, can lead to a significant wave-field back-radiated in the basin. In the following, we will then try to emphasize on the influence of surface structures on seismic site-effects through a simplified numerical model of an alluvial basin located in Nice (French Riviera).

[1]Laboratoire Central des Ponts et Chaussées, Eng. Modelling Dept, 58 bd Lefebvre,
75732 Paris Cedex 15, France, semblat@lcpc.fr
[2]Laboratoire Central des Ponts et Chaussées, Paris, France
[3]LGIT/LCPC, BP 53X, 38041 Grenoble Cedex, France
[4]CETE Méditerranée, 56 bd de Stalingrad, 06300 Nice, France

## Soil-structure-soil interaction (SSSI)

### Experiments for evidences of SSSI

The in situ experiments performed in Volvi Euroseistest in Greece considered a reduced-scale structure at the surface of an alluvial basin very well characterized (Guéguen, Bard, Oliveira 2000). A dynamic loading was applied at the top of the structure and the seismic wavefield radiated in the surface soil layers was measured. The most interesting result is that the amplitude of the radiated wavefield could be large and could remain significant for large propagating distances. Two main parameters govern the soil-structure-soil interaction (SSSI) : the coincidence between soil and structure eigenfrequencies (resonance effect) and the velocity contrast between soil layers (trapped waves effect). Various types of preliminary models were considered to investigate SSSI : semi-analytical, numerical...(Bard 2002)

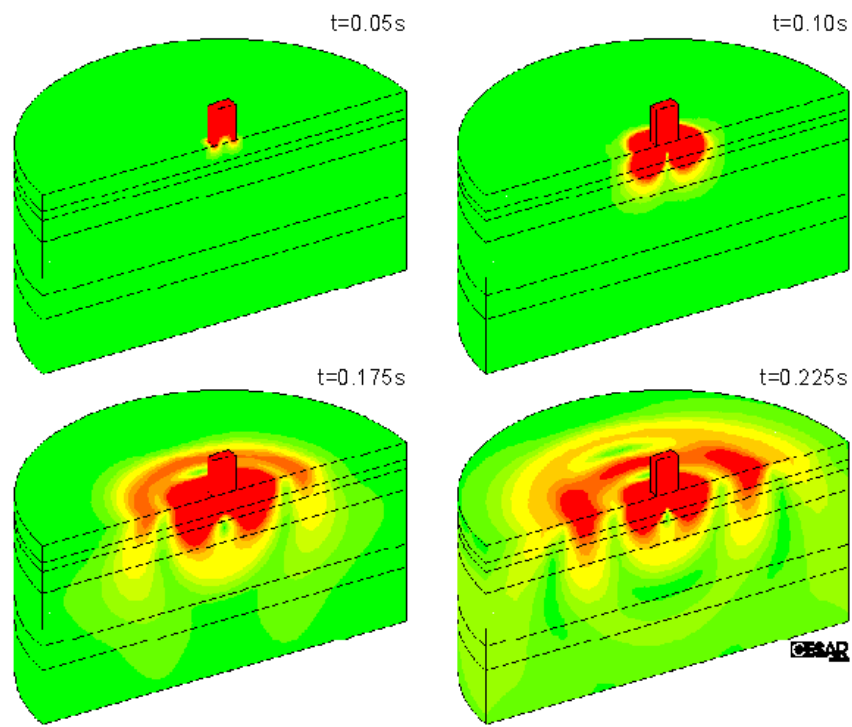

Figure 1.   Modelling of soil-structure-soil interaction from the Volvi Euroseistest experiments.

**Numerical results for SSSI**

Fig. 1 displays the 3D seismic wavefield radiated in the surface soil layers from the Volvi reduced scale structure. The radiated wavefield was computed by the finite element method considering the properties of the different soil layers. Both following features have been studied in details : wavefield directivity and trapped waves effects. As shown by the computations of Fig. 1 (in agreement with the experiments), the motion amplitude remains significantly high during propagation since the amplitude is around 25% of the source amplitude at two times the building base size and still 5% at ten times the building base size (Guéguen, Bard, Oliveira 2000). The effect of directivity is also found to be strong when compared with the various experiments performed (axial and lateral excitations, axial, lateral and diagonal measurements).

## Site-city interaction

**Some experimental evidences and previous analysis**

In the case of multiple structures at the surface of an alluvial deposit, the soil-structure-soil interaction is generalized to multiple interactions and called "site-city interaction" (Guéguen, Bard, Semblat 2000, Bard 2002). For some specific alluvial sites in large cities, there could be a coincidence between eigenfrequencies of surface soil layers and those of tall buildings. The seismic free-field motion could then be influenced by the structural surface vibrations. The effect of the various seismic wavefields radiated from each individual structure can significantly modify the "original" amplification process in the alluvial basin.

In the case of Mexico city, the soil/urban configuration is favourable for site-city interaction to take place. As shown in (Guéguen, Bard, Semblat 2000), energy considerations lead to a large ratio between kinematic energy in the topmost soil layer and the kinematic energy in the building stock. For the city of Nice (French Riviera), such a ratio could reach a large value. Since detailed investigations of site effects in the centre of the town were previously performed (Semblat, Duval, Dangla 2000a), we will analyze the modifications of the free field amplification due to site-city interaction in Nice.

## From site effects to site-city interaction

**Site effects modelling by BEM**

Detailed investigations of site effects in the centre of Nice were performed through various methods : microtremor recordings, real earthquakes measurements, BEM propagative model and simplified modal approach (Semblat, Duval, Dangla 2000a). For a specific alluvial deposit, one considers the amplification factor values given by the Boundary Element Method (Bonnet 1999, Dangla 1988) for several frequencies (Fig. 2). The numerical analysis gives very detailed results on the amplification level, the occuring frequencies, the location... As shown in Fig. 2, for weak motions under shear wave (SH) excitation, the amplification factor is small below 1 Hz and reaches a large value (15) around 1.6 Hz. For larger frequency values, amplification is lower below 2 Hz and increases fastly above this value in the thinnest part of the alluvial deposit. One will analyze the influence of surface structures on the free-field amplification depicted in Fig. 2 (Semblat, Duval, Dangla 2000a).

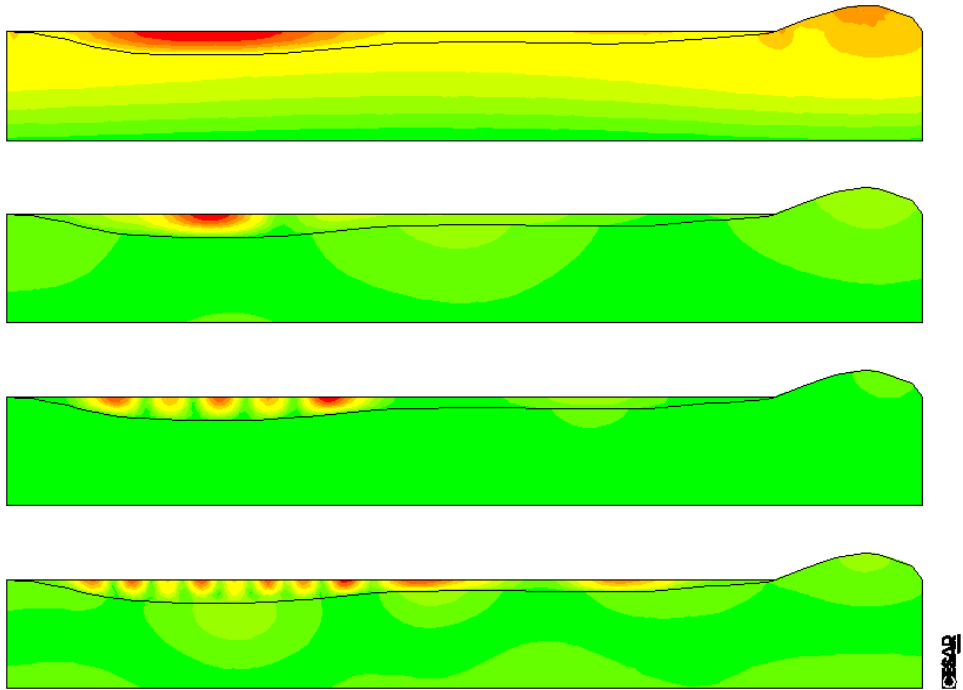

Figure 2. Site effects modelled by BEM in the centre of Nice (French Riviera) : isovalues of the amplification factor $A_i$ at frequency $f_i$ (from top to bottom : $f_1$=0.4 Hz, $A_1$=1.2 ; $f_2$=1.2 Hz, $A_2$=6.5 ; $f_3$=1.6 Hz, $A_3$=15 ; $f_4$=2.0 Hz, $A_4$=6.0).

**BEM model for site-city interaction**

To investigate site-city interaction in the case of Nice city, one considers a BEM model similar to that involved in the analysis of site-effects. For sake of simplicity, a single type of building is considered and several different building densities are chosen. The BEM model used for the analysis presented in Fig. 2 is then combined with other boundary elements describing the surface structures. A specific type of building with an eigenfrequency of 0.6 Hz is firstly considered and five different building densities are taken : 1, 3, 7, 15 and 30 identical structures along the 2 km wide basin. It is not possible to present other cases (various building types, heterogeneous urban configurations) in details in this paper and we will focus on the influence of building density for one type of structure only.

**Modifications due to site-city interactions**

The modifications of site effects due to the various building densities at the surface are shown in Fig. 3 at 1.1 Hz. As shown in the isovalues plots, the buildings have a significant influence on site effects for densities greater or equal to $d_3$. It can be noticed that the motion amplitude is very different from one building to another. Furthermore, for largest densities, the areas of strong amplification are significantly influenced (amplification level, size, location) by the dynamic response of surface structures. One has to precisely quantify the contribution of buildings to the modifications of free-field site effects.

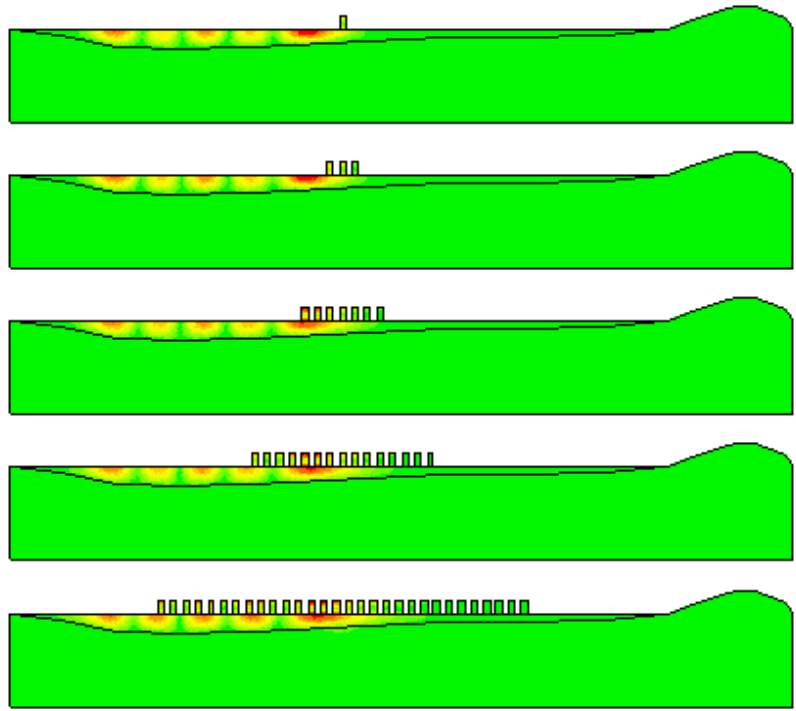

Figure 3.  Site-city interaction in Nice city centre : isovalues of the amplification factor for various building densities.

**Influence on surface motion amplification**

To compare free-field motion amplification (previous site effects estimation) to the present analysis of site-city interaction, the surface amplification is compared in both cases for various building densities at some specific frequencies (Fig. 4). For the first frequency value (0.5 Hz), the influence of buildings for low densities is rather small since there is just a local modification with no significant increase of the overall maximum amplification. For larger densities, the modification of site effects is very important and there is probably a group effect leading to a strong increase of the amplification level (three times) when compared to the free-field case. The location of maximum site effects is also different since it is moved towards the buildings group in eastern direction (right). For such wavelengthes, the contribution of each building could be easily identified in Fig. 4. In the same figure, for the second frequency value (0.6 Hz, i.e. eigenfrequency of the specific building type considered), a clear resonance of the building alone model can be noticed with a very strong increase of the amplitude (from 5 to above 20). The amplification level nevertheless decreases away from the structure (when compared to the free-field amplification). The modifications are significant but much lower for intermediate density values. For the two largest ones, complex multiple resonances and group effects appear with a rather strong increase of the amplification factor.

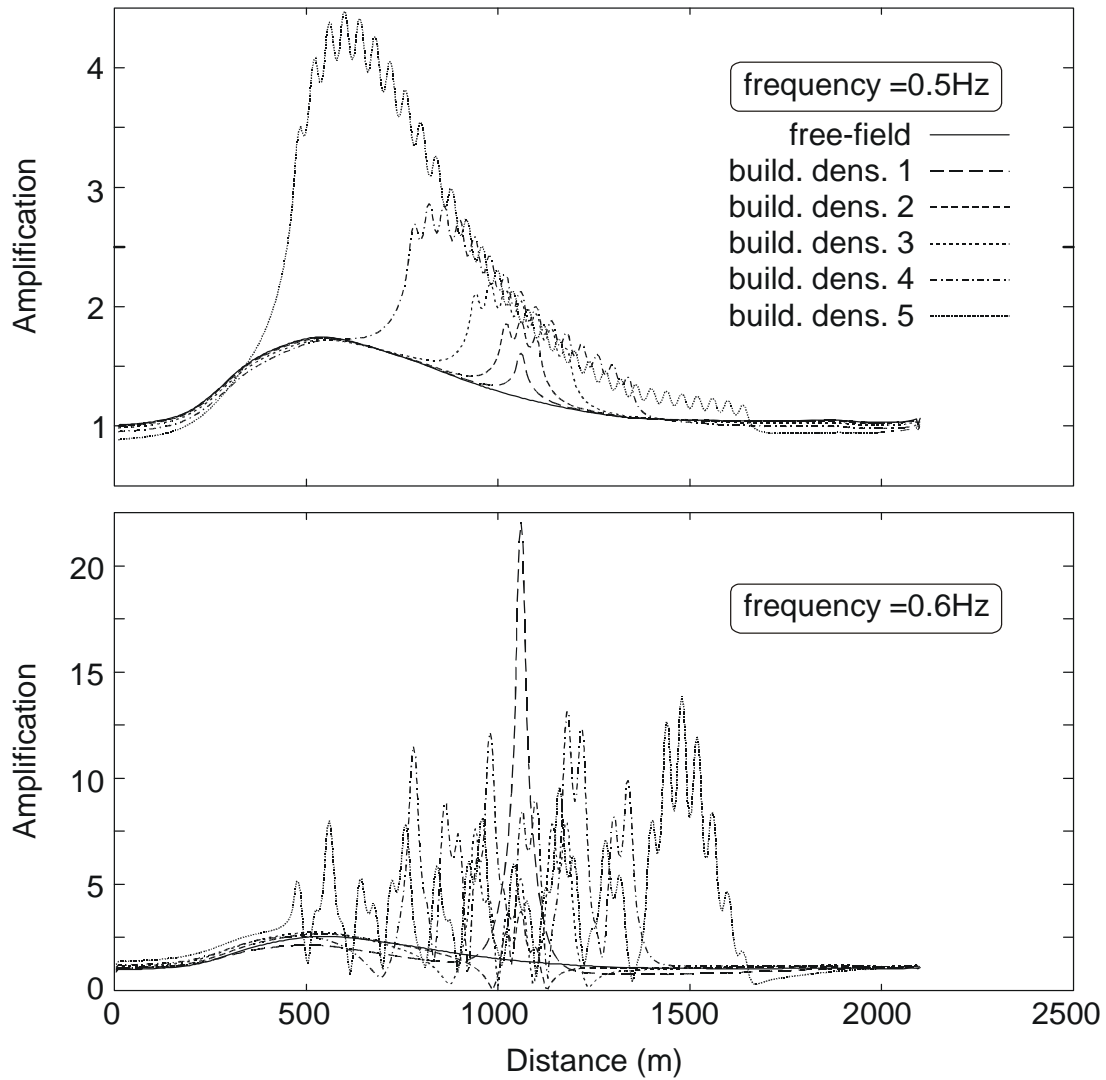

Figure 4.  Influence of site-city interaction on the amplification of surface ground motion (frequencies 0.5 and 0.6Hz).

In Fig. 5, frequency values are above the eigenfrequency of the homogeneous type of building. For the first frequency (1.1 Hz), the form of the amplification curve is partially identical but the amplification level is significantly modified depending on the building density. When compared to the free-field amplification, low urban density cases lead to an increase of the amplification factor between 20 and 50%. For the two largest building densities, the amplification is close or even lower than the free-field values. The reduction can reach 40% in some areas of the basin. For this frequency value and for all densities, the areas of maximum amplification always correspond to the same location. Modifications of the free-field site amplification due to site-city interaction consequently appear as a global increase (or decrease) of the level.

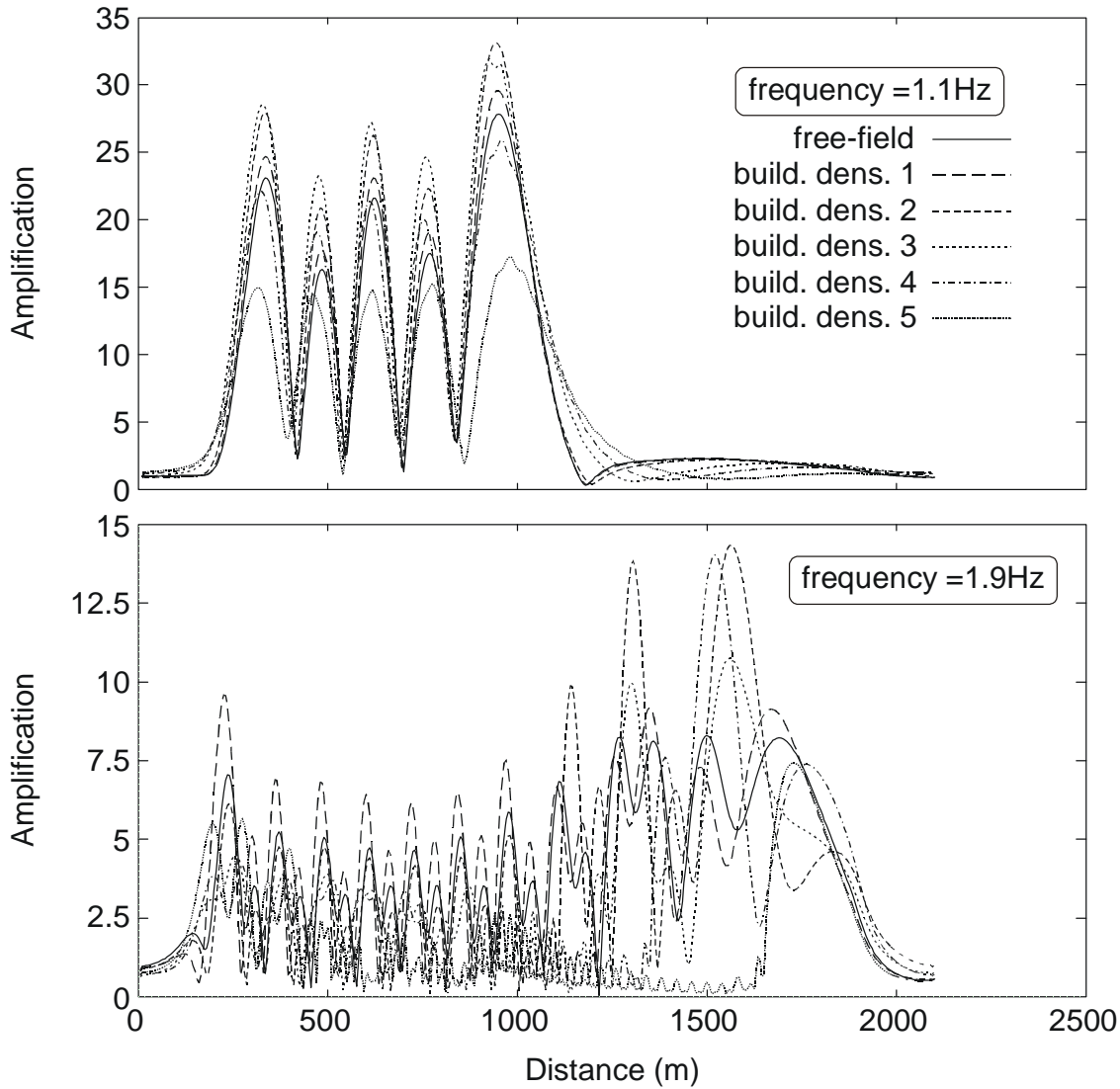

Figure 5.  Influence of site-city interaction on the amplification of surface ground motion.

For the second frequency value given in Fig. 5 (1.9 Hz), surface amplification is lowered in a very large part of the alluvial deposit (West). However, the first density value leads to larger amplification values (30%) than in the free-field case at similar locations. Whereas in the eastern (thinnest) part of the basin, very large increases of the reference amplification can be noticed. The location of maximum values is very different from one density to another. For densities No.2 and 3, we even get a large amplification peak instead of a local minimum value as in the free-field case. The combination of single and multiple structural resonances with complex propagative amplification phenomena then leads to strong modifications of original site effects in the case of a "pure free-field" amplification. The influence of surface structures is generally large on the amplification level and could be very small on the location of the maxima. As shown in Figs. 4 and 5, the respective values of the eigenfrequency of the building type and of the excitation frequency are very important for the potential appearance of strong site-city interaction.

## Conclusion

Starting from soil-structure-soil interaction (wavefield radiated back into the soil from a vibrating structure), site-city interaction has been studied through the modifications of site effects considering multiple radiated wavefields. For the specific site considered in Nice (French Riviera), the original free-field amplification can be strongly modified by site-city interaction with increases up to 50% of the amplification level. Around the resonant frequency of the building type, some particular results are obtained in this homogeneous urban configuration. The influence of site-city interaction can also be significant on the location of maximum amplification areas as shown in the paper for the lowest and largest frequency values. The main issues to consider for future work would be mainly to consider group effects, other types of buildings and heterogeneous urban configurations. Taking into account the computational cost, one also has to investigate more realistic three-dimensional urban models. Beyond the influence of site-city interaction on site effects, one also could analyze the effect of structure to structure through the soil interaction in the case of dense urban configurations.

## Acknowledgments

The authors acknowledge the financial support of project ACI-CATNAT on *" Site-city interaction and seismic hazard in urban areas "* (CNRS/INSU). This work is also part of the research program *" Seismic risks "* of the Lab. Central des Ponts & Chaussées (LCPC, Paris).